\documentclass[prl,floatfix,amsmath,superscriptaddress,singlecolumn]{revtex4}
\usepackage{amssymb}
\usepackage{graphicx}
\usepackage{graphics}
\usepackage{amsmath}
\usepackage{amsthm}
\usepackage{color}
\usepackage{float}

\newcommand{\bea}{\begin{eqnarray}}
\newcommand{\eea}{\end{eqnarray}}
\def\bi{\begin{itemize}}
\def\ei{\end{itemize}}
\def\bc{\begin{center}}
\def\ec{\end{center}}

\def\C{\hbox{$\mit I$\kern-.7em$\mit C$}}
\def\R{\hbox{$\mit I$\kern-.6em$\mit R$}}

\def\ket#1{|#1\rangle}
\newcommand{\one}{\mbox{$1 \hspace{-1.0mm}  {\bf l}$}}

\def\ket#1{\left| #1\right>}
\def\bra#1{\left< #1\right|}

\newtheorem{theorem}{Theorem}

\begin{document}
\author{C. Spee}
\affiliation{Institute for Theoretical Physics, University of
Innsbruck, Innsbruck, Austria}
\author{J. I. de Vicente}
\affiliation{Departamento de Matem\'aticas, Universidad Carlos III de
Madrid, Legan\'es (Madrid), Spain}
\author{B. Kraus}
\affiliation{Institute for Theoretical Physics, University of
Innsbruck, Innsbruck, Austria}
\title{Few-body entanglement manipulation}

\begin{abstract}
In order to cope with the fact that there exists no single maximally entangled state (up to local unitaries) in the multipartite setting, we introduced in \cite{MESus} the maximally entangled set of $n$-partite quantum states. This set consists of the states that are most useful under conversion of pure states via Local Operations assisted by Classical Communication (LOCC).
We will review our results here on the maximally entangled set of three- and generic four-qubit states. Moreover, we will discuss the preparation of arbitrary (pure or mixed) states via deterministic LOCC transformations. In particular, we will consider the deterministic preparation of arbitrary three-qubit (four-qubit) states via LOCC using as a resource a six-qubit (23-qubit) state respectively.
\end{abstract}
\maketitle

\section{Introduction}

Bell's theorem states that the predictions of quantum mechanics are not compatible with any local realistic theory (i.\ e.\ any local hidden variable model) \cite{bell}. This result has profound conceptual consequences and has shaped our understanding of quantum mechanics. However, 50 years after its discovery, the relevance of this result goes beyond foundational issues. The development of quantum information theory \cite{nielsenandchuang} in the last decades has taught us that the counter-intuitiveness of quantum mechanics can be exploited to devise technologies beyond what can be reached classically. Interestingly, it is the correlations present in entangled quantum states that allow to violate Bell inequalities and it is entanglement what is considered to be the fundamental ingredient behind many of the applications of quantum information processing. Nowadays, entanglement is regarded as a resource and a big effort has been put up in the last twenty years to develop a theory of entanglement, which aims at its characterization, manipulation and quantification. Central to this theory is the paradigm of Local Operations assisted with Classical Communication (LOCC). These are the most general operations allowed by the rules of quantum mechanics to spatially separated parties: each subsystem can undergo any possible form of quantum dynamics at will and the parties can correlate these local protocols via classical communication. The restriction to LOCC is, thus, a very natural one and at the heart of many applications. Moreover, as entanglement cannot be created by LOCC alone, it can be considered to be a resource to overcome the limitations of state manipulation restricted to this class of operations.

Unfortunately, the set of LOCC transformations is mathematically subtle and to characterize the possible conversions among entangled states is in general a very hard problem. Entanglement theory is much better established in the bipartite case than in the multipartite one. This is partly because  entanglement manipulation via LOCC is better understood in the former setting. This has allowed to identify the bipartite maximally entangled state as the most useful state under LOCC transformations, to introduce the scenario of entanglement distillation and to define several operational entanglement measures \cite{horodecki}. On the other hand, not surprisingly, most bipartite quantum information protocols such as teleportation \cite{teleport} and cryptography schemes \cite{crypto} are optimally implemented by means of the bipartite maximally entangled state. Although several applications are also known for multipartite entanglement (e.g. quantum secret sharing \cite{secretsharing} and one-way quantum computation \cite{one-way}), a more systematic and profound study of LOCC manipulation in this realm appears to be of paramount importance. In addition to playing a pivotal role in the theory of entanglement, this would identify the most useful multipartite states and could lead to new applications of quantum information in many-body scenarios.

In this paper we review our recent results on the characterization of LOCC transformations over few-body pure states \cite{MESus}. As a unique maximally entangled state does not exist in this situation, we introduced the notion of the maximally entangled set of states, as those that are most useful under LOCC conversions. We characterized the maximally entangled set for systems of three and four qubits. In addition to this, we present LOCC protocols that allow to obtain arbitrary (pure or mixed) bipartite, three or four-qubit states from a single multipartite state with more subsystems. In particular, we provide a specific six-qubit state that allows to prepare any three-qubit state by LOCC and the analogous for any four-qubit state with a given 23-qubit state.
\section{Preliminaries}
Entanglement is well understood in the bipartite pure state case. In this case it  is well known that any state can be written in Schmidt decomposition \cite{nielsenandchuang}, i.e. it can be written as $U_A\otimes U_B\sum_{i=1}^d \sqrt{\lambda_i} \ket{ii}$, where $U_A$ and $U_B$ are Local Unitaries (LUs), $d$ is the dimension of the smaller subsystem, $\lambda_i\geq 0$, $\sum_i \lambda_i=1$ and $\ket{i}$ are computational basis states. The coefficients $\sqrt{\lambda_i}$ are called Schmidt coefficients. The entanglement properties of a bipartite pure state are completely characterized by its Schmidt coefficients, as states that are equivalent up to LUs have the same entanglement. In order to see this consider two states, $\ket{\Psi}$ and $\ket{\Phi}$, that are LU-equivalent, i.e. $\ket{\Psi}=V_1\otimes \ldots \otimes V_n\ket{\Phi}$. By applying LUs (which corresponds to a deterministic LOCC protocol) one can convert $\ket{\Psi}$ to $\ket{\Phi}$ and vice versa. As entanglement is non-increasing under LOCC, one obtains that $E(\Phi)\geq E(\Psi)$ and $E(\Phi)\leq E(\Psi)$ for any entanglement measure $E$, i.e. these states have the same entanglement ($E(\Phi)= E(\Psi)$).  Interestingly, the converse is also true. It has been shown that if one can transform a pure state $\ket{\psi}$ via LOCC deterministically to a pure state $\ket{\phi}$ and vice versa, then the two states are LU-equivalent \cite{circleLU}. As applying LUs does not change the entanglement of a state, the entanglement properties of a bipartite pure state are completely determined by the Schmidt coefficients. Moreover, in this case it can be shown that any LOCC protocol can be simulated by a (simple) LOCC transformation where one party performs a POVM measurement and the other party applies depending on the outcome a LU \cite{LOCCbipart}. As only one-way communication is necessary this simplifies the study of LOCC transformations and entanglement for the bipartite case. In fact, LOCC transformations among bipartite pure states have been characterized completely by Nielsen \cite{nielsen}. In particular, it has been shown in \cite{nielsen} that a state $\ket{\psi}$ can be transformed via LOCC deterministically to state $\ket{\phi}$ iff $\vec{\lambda}_\phi\succ\vec{\lambda}_\psi$, i.e. $\vec{\lambda}_\phi$ majorizes $\vec{\lambda}_\psi$. The vectors $\vec{\lambda}_{\phi(\psi)}$ correspond to $d$-dimensional vectors containing the squares of the Schmidt coefficients of $\ket{\phi}(\ket{\psi})$ respectively. A vector $\vec{y}=(y_1,\ldots, y_d)$ is said to majorize $\vec{x}=(x_1,\ldots, x_d)$ (i.e.  $\vec{y}\succ\vec{x}$)  if $\forall k\in\{1,\ldots,d-1\}$\bea\sum_{i=1}^kx_i^\downarrow\leq\sum_{i=1}^ky_i^\downarrow\eea
and \bea\sum_{i=1}^dx_i=\sum_{i=1}^dy_i.\eea
Note that $\downarrow$ denotes that the components are ordered in non-increasing order (e.g. $x_d^\downarrow$ is the smallest and $x_1^\downarrow$ is the largest component of $\vec{x}$). Note further that in the two qubit case this result implies that LOCC imposes a total order on the states with respect to their entanglement. In the case of two $d$-dimensional systems with $d> 2$ the situation changes, as  there are states that are incomparable. More precisely, there exist states, $\ket{\Psi}$ and $\ket{\Phi}$, for which it holds that one can neither transform  $\ket{\Psi}$ into $\ket{\Phi}$ via deterministic LOCC transformations , i.e. $\vec{\lambda}_\Phi\not\succ\vec{\lambda}_\Psi$ nor $\ket{\Phi}$ into $\ket{\Psi}$, i.e. $\vec{\lambda}_\Psi\not\succ\vec{\lambda}_\Phi$. This shows a clear difference between the two-qubit case and the case of two $d$-dimensional systems with $d> 2$. Nevertheless, in both cases there exists a single  state (up to LUs) that allows to obtain any other state via deterministic LOCC transformations, namely the state $\ket{\Phi^+}_d=1/\sqrt{d} \sum_{i=1}^d \ket{ii}$.  

As knowing which LOCC transformations are possible allows to impose a (partial) order on the states with respect to their entanglement it allows to identify entanglement measures. To be more precise, the condition that an entanglement measure $E$ has to fulfill is that it is non-increasing under LOCC. Therefore, any function $E$ (defined on bipartite pure states) that respects the ordering due to the majorization condition, i.e. for $\vec{\lambda}_\phi\succ\vec{\lambda}_\psi$ it holds that $E(\psi)\geq E(\phi)$, is an entanglement measure for pure states. Moreover, using the majorization condition it is easy to identify the maximally entangled states. As entanglement is non-increasing under LOCC, a maximally entangled state can not be obtained via LOCC (excluding LUs) deterministically from any other state. Thus, the condition is that its Schmidt vector does not majorize any other Schmidt vector. It is easy to see that up to LUs this property is fulfilled by a single state, namely  $\ket{\Phi^+}_d=1/\sqrt{d} \sum_{i=1}^d \ket{ii}$. Note that any other state can be reached via LOCC from this state, as its Schmidt vector  is majorized by any other Schmidt vector. Hence, this state is the most useful one concerning applications. If a protocol is based on another bipartite state with smallest dimension of the subsystems smaller than or equal to $d$, it can be also implemented using $\ket{\Phi^+}_d$ as a resource by applying the corresponding LOCC protocol beforehand. On the other hand, if a protocol requires that the parties share a maximally entangled state, then there exists no other (LU-inequivalent) state  that allows to perform the same task. For example, faithful teleportation \cite{teleport}  requires  a  maximally entangled state as a resource.

Characterizing LOCC transformations in the multipartite setting is hard, as one has to deal with several rounds of communication. It has even been proven that some tasks require infinitely many rounds \cite{infLOCC}. Moreover, LOCC is not closed, i. e. there exist sequences of LOCC protocols $\{\Lambda_1, \Lambda_2,\ldots\}$ such that $\lim_{n\rightarrow \infty}\Lambda_n$ is not a LOCC transformation \cite{LOCCnotclosed}. 

Due to these difficulties multipartite LOCC transformations have only been characterized for a few classes of states \cite{WGHZLOCC, MESus}. Other classifications of multipartite states with respect to entanglement have been studied such as LU-equivalence \cite{LUba} and SLOCC-equivalence \cite{3qubitSLOCC,4qubitSLOCC}. As already argued before, states that are in the same LU-equivalence class have the same entanglement properties. The problem of deciding whether two pure $n$-qubit states are LU-equivalent has been solved in \cite{LUba}. By applying LUs a generic state can be brought into its unique standard form. Two generic states are then LU-equivalent iff their standard forms are the same. For non-generic states there exists an algorithm that allows to decide whether two states are LU-equivalent or not \cite{LUba}. Another classification can be established by considering Stochastic LOCC (SLOCC) transformations. Two states, $\ket{\Psi}$ and $\ket{\Phi}$, are said to be in the same SLOCC class, if one can transform $\ket{\Psi}$ via LOCC with non-zero probability into $\ket{\Phi}$ and vice versa.  Mathematically, this implies that $\ket{\Psi}$ can be written as $g\ket{\Phi}$, where $g\in\mathcal{G}$ with $\mathcal{G}$ being the set of local invertible operators, i.e. $g=g^1\otimes\ldots\otimes g^n$ and $g^i\in GL(2)$ \cite{3qubitSLOCC}. Note that in contrast to deterministic LOCC transformations the classification according to SLOCC (or LU) corresponds to an equivalence relation. States in different SLOCC classes have different kinds of entanglement but SLOCC does not impose any order with respect to their entanglement. For three qubits there are two different truly tripartite entangled SLOCC classes, the W-class and the GHZ-class \cite{3qubitSLOCC}. As representatives for the GHZ- and the W-class one can choose the GHZ-state \cite{ghz}, i.e. $1/\sqrt{2}(\ket{000}+\ket{111})$ and the W-state \cite{3qubitSLOCC}, i.e. $1/\sqrt{3}(\ket{100}+\ket{010}+\ket{001})$ respectively. In the four-qubit case it has been shown that there exist infinitely many SLOCC classes which can be grouped into 9 different families \cite{4qubitSLOCC}. 

As the mathematical study of multipartite LOCC transformations is formidably hard, other approaches towards the characterization of entanglement  have been pursued.  They consist of enlarging the set of allowed operations such that one obtains classes of operations that are easier to deal with mathematically as for example  PPT preserving maps \cite{ppt} or separable transformations (SEP) (see \cite{gour} and references therein). As it will be important for our discussion we will focus here on SEP. LOCC is strictly included in SEP \cite{sep}, i.e. any LOCC transformation corresponds to a completely positive trace-preserving map $\Lambda$  whose action on any density matrix $\rho$ can be written as $\Lambda (\rho) =\sum_i X_i \rho X_i^\dagger $ where $X_i= x_i^{(1)}\otimes x_i^{(2)}\ldots\otimes x_i^{(n)}$, $ x_i^{(j)}$ is a complex matrix and  $\sum_i X_i^\dagger X_i=\one$. However, there exist SEP transformations that can not be implemented via LOCC deterministically \cite{sepnotLOCC}. Note that in order for $\Lambda$ to correspond to a LOCC protocol the operators $X_i$ have to fulfill some condition apart from being local, i.e. each $X_i$ denotes the operation corresponding to one branch of the LOCC protocol. To be more precise, $ x_i^{(j)}$ is a product of operators where each of these operators originates from the implementation of a POVM and in general there is a dependence between the implemented POVM and previous measurement outcomes.  Obviously, SEP transformations are mathematically  much easier to characterize. Unfortunately, however, they lack  an operational meaning. Nevertheless, the result of \cite{gour} on convertibility via SEP transformations has been crucial for determining the maximally entangled set for three and four qubits as we will explain in the next section.

\section{The maximally entangled set}
As argued in the previous section, the entanglement properties of multipartite states are hard to characterize. This is also reflected in the fact that there exist several different notions of maximally entangled multipartite states in the literature (see e.g. \cite{bipartmax, averagemax}). Why is it of importance to know which states are maximally entangled? Apart from fundamental interest,
these states are the most useful ones concerning application. Any protocol that might rely on the parties sharing a state that is not maximally entangled can also be performed using as resource a maximally entangled state (by applying the LOCC protocol beforehand that transforms the corresponding maximally entangled state deterministically into the state that the protocol is based on), but the converse is not true. Moreover, these states are promising candidates to discover new applications for multipartite entanglement, as also most bipartite applications like teleportation \cite{teleport} and cryptography \cite{crypto} rely on the parties sharing a maximally entangled state. In order to characterize maximal entanglement one has to develop a deeper understanding of multipartite LOCC transformations. Knowing which LOCC transformations are possible allows to define new (operational) entanglement measures. This is due to the fact that, as mentioned before, the condition an entanglement measure has to obey is that it is non-increasing under deterministic LOCC transformations.

As already discussed,  it is well known which states are maximally entangled in the bipartite case. They are those states which correspond up to LUs to $\ket{\Phi^+}_d\propto \sum_{i=1}^d \ket{ii}$, where $d$ denotes the dimension of the smaller subsystem. As the application of LUs does not change the entanglement properties of a state and therefore two states in the same LU-equivalence class are equally useful, we will in the following only consider one representative per LU-equivalence class. Hence, $\ket{\Phi^+}_d$ is \textit{the} maximally entangled state.  Recall that the important property of $\ket{\Phi^+}_d$ (and the reason why it is maximally entangled) is that it can not be reached via deterministic LOCC transformations from any other state (excluding LU) and that any state can be obtained deterministically via LOCC from this state \cite{nielsen}.

In the multipartite setting there exists no single state (up to LUs) with this property. This can already easily be seen in the three-qubit case. Recall that there are two different truly three-partite entangled SLOCC classes, the W-class and the GHZ-class. If there would exist a state $\ket{\Psi}$ that can be transformed via a deterministic LOCC transformation to a state in the GHZ-class $\ket{\Psi_{GHZ}}$ as well as to a state in the W-class $\ket{\Psi_{W}}$, then it would hold that there exist complex 2 x 2 matrices  $ M_k^i, N_k^i$ such that $M_k^1\otimes M_k^2\otimes M_k^3\ket{\Psi}=\ket{\Psi_{GHZ}}$ and $N_k^1\otimes N_k^2\otimes N_k^3\ket{\Psi}=\ket{\Psi_{W}}$. As we only consider transformations between truly tripartite entangled states this implies that the matrices $M_k^i$ and $N_k^i$ have to be invertible for all $i$ and therefore $\ket{\Psi_{GHZ}}$ and $\ket{\Psi_{W}}$ would be in the same SLOCC class \footnote{Note that this line of argumentation can be easily generalized to the $n$-qubit case. Thus, deterministic LOCC transformations among fully entangled $n$-qubit states are only possible between states in the same SLOCC class.}. Thus, deterministic LOCC transformations among fully-entangled three-qubit state are only possible between states in the same SLOCC class and so there exists no single three-qubit state that allows to reach any other state via deterministic LOCC transformations. 

In order to cope with the fact that there exists no single maximally entangled state in the multipartite setting, we introduced in \cite{MESus} the Maximally Entangled Set (MES) of $n$-partite states.
The MES of $n$-partite states, $MES_n$, is defined as the minimal set of $n$-partite pure states such that any other truly $n$-partite entangled pure state can be reached via LOCC deterministically from one of the states in $MES_n$ \footnote{Note that $MES_n$ is unique (up to LUs).}. To state it differently, $MES_n$ is the set of $n$-partite states characterized by the following two properties:
\begin{enumerate}
\item No state in $MES_n$ can be reached via LOCC (excluding LU) deterministically from any other $n$-partite state.
\item For any truly $n$-partite entangled state $\ket{\Psi}\not\in MES_n$ there exists a state $\ket{\Phi}\in MES_n$ such that $\ket{\Psi}$ can be obtained via a deterministic LOCC transformation from $\ket{\Phi}$.
\end{enumerate}
Hence, $MES_n$ is the minimal set of states that are most useful with respect to any application. As already explained before, any application that uses a state that is not in the MES can also be performed by using a state that is in the MES (but not vice versa). Note that the MES for bipartite $d$-level systems is given by $\{\ket{\Phi^+}_d\}$. 

Another concept that is important for our study is the one of isolation \cite{MESus}. Isolated states are defined as truly $n$-partite entangled states that can neither be reached nor can they be converted to any other truly $n$-partite entangled pure state via LOCC (excluding LU) in a deterministic way. It is obvious from this definition that isolated states have to be contained in the MES. The subset of the MES of LOCC convertible states is of particular interest, as it is the only relevant set of states concerning deterministic entanglement manipulation and therefore will most probably play an important role in discovering new applications of multipartite entanglement.

\begin{figure}[H]
\begin{center}
  \includegraphics[scale=0.45]{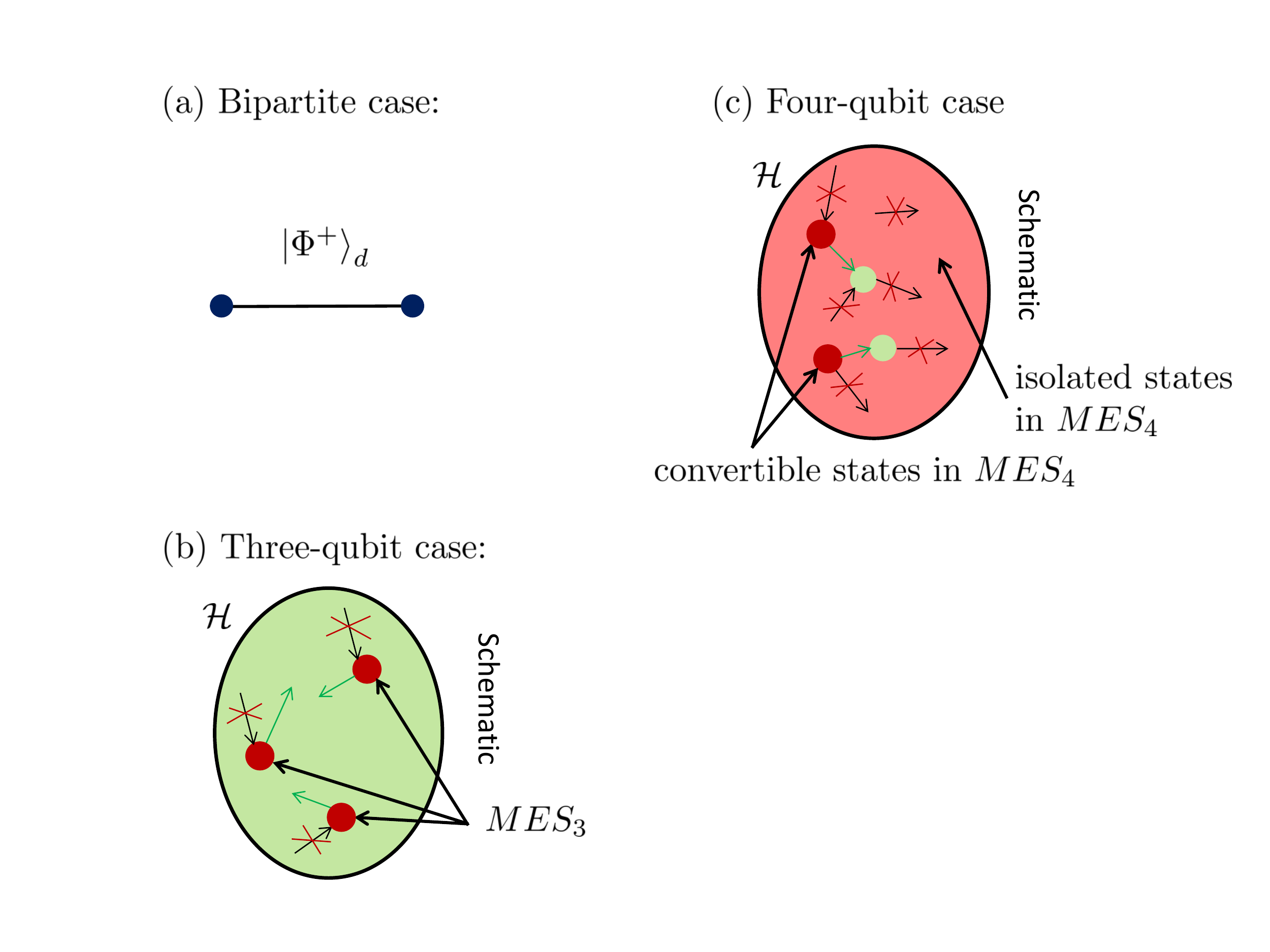}
  \caption{This graphic shows schematically the MES for the bipartite-, three- and four-qubit case. In (b) and (c) the reachable states are indicated in green, the isolated states in the MES in light red and the convertible states in the MES in dark red.}
\label{graphMES}
\end{center}
\end{figure}

In \cite{MESus} we determined the MES for three-qubit and generic four-qubit states. In contrast to the two-qubit case where the MES is given by $\{\ket{\Phi^+}_2\}$ the MES for three qubits, $MES_3$, contains infinitely many states.  It is characterized by 3 parameters, whereas an arbitrary three-qubit state is characterized by 5 parameters  (up to LUs) \cite{5par, 3qubitLU}. Therefore, $MES_3$ is of measure zero.  Interestingly, no state in $MES_3$ is isolated (see Fig. \ref{graphMES}). The picture changes again drastically when going from the three-qubit to the four-qubit case. For four qubits $MES_4$ is of full measure. This is due to the fact that almost all states are isolated, i.e. deterministic LOCC transformations are hardly ever possible among fully entangled four-qubit states. Interestingly, the subset of non-isolated states in the MES is of measure zero (see Fig.  \ref{graphMES}). As already mentioned, these states are the only relevant ones for entanglement manipulation.

Let us now present our results on $MES_3$, the MES of three qubits, in more detail. Due to the existence of  two truly tripartite-entangled SLOCC classes for three qubits \cite{3qubitSLOCC}, the W-class and the GHZ-class,
 $MES_3$ has to contain at least two states. Recall that if  $\ket{\Psi}$ and $\ket{\Phi}$ are in the same SLOCC class, then $\ket{\Psi}$ can be written as $g\ket{\Phi}$, where $g=g^1\otimes\ldots\otimes g^n$ and $g^i\in GL(2)$. Thus, we can write any state in the GHZ-class as $g\ket{GHZ}$. Note that the local operator $g$ is not unique, as the two states $g\ket{GHZ}$ and $(g S)\ket{GHZ}$ coincide if S is a local symmetry of the GHZ-state. Here and in the following, we denote by $S(\Psi)=\{S\in\mathcal{G}:S\ket{\Psi}=\ket{\Psi}\}$ the local symmetry group of the state $\ket{\Psi}$. In order to get rid of this ambiguity, we defined in \cite{MESus} a standard form for states in the GHZ-class. In particular, we showed that any state in the GHZ-class can be written up to LUs as
\bea\label{standformghz}
g_x^1\otimes g_x^2\otimes g_x^3 P_z\ket{GHZ},
\eea
where $P_z=diag(z,1/z)$ with $z\in\C \backslash 0$ and $g_x^i=\sqrt{G_x^i}$ with $G_x^i=(g_x^i)^\dagger g_x^i= 1/2\one + \gamma_x^i \sigma_x$ and $0\leq\gamma_x^i<1/2$. Here and in the following we denote by $\one$ and $\sigma_w$ where $w\in\{x,y,z\}$ the identity operator and the Pauli matrices.  Note that the restriction $\gamma_x^i\in [0,1/2)$ ensures that the operators $g_x^i$ are invertible, as otherwise entanglement would be destroyed.  

It is well known that any state in the W-class can be written (up to LUs) as $g_1\otimes g_2\otimes \one\ket{W}$ \cite{3qubitSLOCC}, where $\ket{W}\propto \ket{001}+\ket{010}+\ket{100}$,
\bea \label{Wmatrices} g_1=\begin{pmatrix} 1&0\\ 0& x_1/x_3
\end{pmatrix}\,\,\,\,\,\ \textrm{and} \quad
  g_2=\begin{pmatrix} x_3&x_0\\ 0& x_2 \end{pmatrix},\eea
with $x_1, x_2, x_3>0$ and $x_0\geq 0$, i.e. $x_0\ket{000}+x_1\ket{100}+x_2\ket{010}+x_3\ket{001}$.
Using this notation we can now present the MES for three qubits \cite{MESus}.
 \begin{theorem} \label{theo3qubit} The MES of three qubits, $MES_3$, is given by \bea \label{mes3} MES_3=\{g_x^1\otimes g_x^2\otimes
g_x^3 P_z\ket{GHZ}, g_1 \otimes g_2 \otimes \one \ket{W}\},\eea where $z\in\{1,i\}$, no $g^i_x\propto \one$ (except for the GHZ-state) and $g_1$ and $g_2$ are diagonal.
\end{theorem}
The general idea of how to obtain this result will be presented below. Due to Theorem \ref{theo3qubit} a state in the GHZ-class is in $MES_3$ iff it is either the GHZ-state or in its corresponding standard form $z\in\{1,i\}$ and $\gamma_x^i\neq 0$ $\forall i$. States in the W-class which are in $MES_3$ have the property that $x_0=0$ (see Eq. (\ref{Wmatrices})), i.e. their standard form corresponds to $x_1\ket{100}+x_2\ket{010}+x_3\ket{001}$. All three-qubit states that are not in $MES_3$ can be reached via a LOCC protocol deterministically from some state in $MES_3$.
Interestingly, the states in the set $MES_3$ have a simple description in terms of the decomposition presented in \cite{3qubitLU}. Any state in $MES_3$ can be written up to LUs as
\bea\label{3qubitdecomp}\ket{\Psi}=
\ket{0}\ket{\Psi_s}+\ket{1}Y(\beta')\otimes
Y(\beta)\ket{\Psi_s}\},\eea where
$\ket{\Psi_s}=a\ket{00}+\sqrt{1-a^2}\ket{11}$ and $a,\beta, \beta^\prime \in \R$.
In this decomposition it is easy to see that $MES_3$ is characterized by 3 parameters. As three-qubit states up to LUs are characterized by 5 parameters (see Eq. (\ref{standformghz}) and e.g. \cite{5par, 3qubitLU}), $MES_3$ is of measure zero  (see Fig. \ref{graphMES}).
Moreover, one can easily show (by constructing the corresponding LOCC protocol) that all states in $MES_3$ are non-isolated.

Let us proceed by presenting the results on the MES for generic four-qubits states \footnote{Note that for the non-generic cases similar results can be obtained \cite{MESnongeneric}.}\cite{MESus}. A generic four-qubit state belongs to one of the SLOCC classes denoted by $G_{abcd}$ in \cite{4qubitSLOCC}. Its representatives can be chosen to be
\begin{align}
|\Psi\rangle&=\frac{a+d}{2}(|0000\rangle+|1111\rangle)+\frac{a-d}{2}(|0011\rangle+|1100\rangle)\nonumber\\
&+\frac{b+c}{2}(|0101\rangle+|1010\rangle)+\frac{b-c}{2}(|0110\rangle+|1001\rangle),\label{seed}
\end{align}
where $a,b,c,d \in \C$ with  $b^2\neq c^2\neq d^2\neq b^2$, $a^2\neq b^2, c^2, d^2$ and the parameters fulfill the condition that there exists no $q\in \C\backslash 1$ such that $\{a^2,b^2,c^2,d^2\}=\{q a^2,q b^2,q c^2,q d^2\}$. The non-trivial symmetries of these states are given by $\sigma_w^{\otimes 4}$ where $w\in\{x,y,z\}$. As for the three-qubit case one can define a unique standard form (by sorting the parameters in Eq. (\ref{seed}), using the symmetries to uniquely define $g^\dagger g$ and fixing the LUs). In the following we use the notation $g_w^i=\sqrt{G_w^i}$ where $w\in\{x,y,z\}$
and $G_w^i=(g_w^i)^\dagger g_w^i= 1/2\one + \gamma_w^i \sigma_w$ with $|\gamma_w^i|<1/2$. Using this notation we can state our result on which fully-entangled four-qubit states can be reached via a deterministic LOCC transformation from some other state.
\begin{theorem} \label{Thgen4} A generic state, $h \ket{\Psi}$,
is reachable via LOCC from some other LU-inequivalent state iff (up to permutations)
 either $h= h^1\otimes h_w^2 \otimes h_w^3 \otimes h_w^4$, for
$w\in\{x,y,z\}$ where $h^1\neq h_w^1$ or $h= h^1\otimes\one^{\otimes 3}$ with $h^1\not \propto \one$ arbitrary.
\end{theorem}
Note that whereas the reachable states are characterized by 12 parameters a generic state is (up to LUs) described by 18 parameters. Hence, the set of reachable states is of measure zero. This implies that the MES for four qubits is of full measure. That is almost all states are in $MES_4$. By investigating which states are convertible via LOCC it becomes apparent that almost all states are isolated. In the following theorem we state which states are convertible via deterministic LOCC transformations to some other state.
\begin{theorem} \label{theoremISO}
A generic state $g \ket{\Psi}$ is convertible via LOCC to some other LU-inequivalent state iff (up to permutations) $g=g^1\otimes g_w^2
\otimes g_w^3 \otimes g_w^4$ with $w\in \{x,y,z\}$ and $g^1$ arbitrary.
\end{theorem}
Combining Theorem \ref{Thgen4} and \ref{theoremISO}, one obtains that the non-isolated states in $MES_4$ are given by \bea g_w^1\otimes g_w^2
\otimes g_w^3 \otimes g_w^4\ket{\Psi},\eea where $w\in \{x,y,z\}$ and excluding the case where $\gamma_w^i\neq 0$ for exactly one $i$. Note that as these states are characterized by 10 parameters, the subset of LOCC-convertible states in $MES_4$ is of measure zero. These states are particularly interesting as they are in the MES and can be converted deterministically to some other states. Thus, investigating these states could lead to the discovery of new applications.
We will proceed by presenting the general idea of how to prove which states are reachable (or convertible) via determininistic LOCC transformations. As already pointed out, it is very hard to characterize LOCC transformations in the multipartite setting. In fact, in the four-qubit case a classification of LOCC protocols is not known. Nevertheless, the characterization of MES can be achieved as follows. The fact that LOCC protocols are strictly included in SEP \cite{sep} implies that if a state is not reachable (or convertible) deterministically via SEP it is also not reachable (or convertible) deterministically via LOCC. Hence, a state that is not reachable via SEP has to be in the MES. In order to characterize the states that can not be reached (or converted) via SEP we used the result of \cite{gour} where necessary and sufficient conditions for the deterministic convertibility of pure states via SEP have been derived. With this, we could characterize all three- and generic four-qubit states that are not reachable (or convertible) via LOCC. Moreover, we showed that all other states can be reached (or converted) via LOCC by constructing the corresponding LOCC protocols. Interestingly, these protocols turned out to be very simple. Any reachable three- or generic four qubit state can be obtained from some other state via a LOCC protocol were just one party performs a non-trivial measurement and the other parties apply, depending on the measurement outcome, a LU. The detailed proof of which states are in the MES in the three-qubit and generic four-qubit case can be found in \cite{MESus}.

We will in the following present the result of \cite{gour} and discuss some technical details. In \cite{gour} it has been shown that a state $\ket{\Psi_1} = g
\ket{\Psi}$ can be transformed via SEP to $\ket{\Psi_2}= h
\ket{\Psi}$ iff there exists a $m \in N$ and a set of probabilities,
$\{p_k\}_{1}^m$ ($p_k\geq 0, \sum_{k=1}^m p_k=1$) and local symmetries, $S_k \in
S(\Psi)$ such that \bea \label{EqSep} \sum_k p_k S_k^\dagger H S_k=
r G.\eea Here, we use the notation $H=h^\dagger h$,
$G=g^\dagger g$  and
$r=n_{\Psi_2}/n_{\Psi_1}$ with
$n_{\Psi_i}=||\ket{\Psi_i}||^2$. The POVM elements that allow to do this transformation are of the form $M_k=\frac{ \sqrt{p_k}}{\sqrt{r}} h S_k
g^{-1}$. Note that $M_k$, $H$, $G$  and $S_k$ are local operators. From Eq. (\ref{EqSep}) it becomes apparent that the symmetries of the chosen representative of the SLOCC class play an important role for the convertibility via deterministic SEP transformations. To be more precise, deterministic SEP transformations only become possible if there exists at least one non-trivial symmetry and the symmetries specify which transformations are possible. Thus, in order to characterize SEP convertibility it is crucial to know the symmetries of a representative of the corresponding SLOCC class. Moreover, it is important to note that $H$ and $G$ are not uniquely defined by the states, as the state $g\ket{\Psi}$ is the same as $\tilde{g}\ket{\Psi}=g S\ket{\Psi}$ for $S\in S(\Psi)$. Thus, $G$ and $\tilde{G}=S^\dagger G S$ correspond to the same state. This makes it more complicated to characterize which SEP transformations are possible as the only information about the initial and the final state that enters  Eq. (\ref{EqSep}) is (apart from the normalization factor)  $H$ and $G$.  In order to get rid of this ambiguity, we defined for each SLOCC class a standard form, which also takes into account that we only consider one representative per LU-equivalence class.  In order to get a unique correspondence (up to LUs) between the operator $G$ and the state $g\ket{\Psi}$, we choose symmetries $S_i\in S(\Psi)$ such that $\tilde{G}=S_i^\dagger G S_i$ is of a specific form which ensures that there exists no symmetry $S_j\in S(\Psi)$ such that $S_j^\dagger\tilde{G}S_j$ is also of this form. Obviously, it is sufficient to consider only transformations between states in standard form. As outlined before, using this standard form and Eq. (\ref{EqSep}) allowed us to characterize the states that can not be reached (or converted) via LOCC.

Up to now, we discussed deterministic LOCC transformations among pure states. In the following section we will extend our study to the preparation of mixed states.
\section{Deterministic state preparation of an arbitary two-, three- and four qubit state (pure or mixed)}
In this section, we will show that in the bipartite- and three-qubit case having access to all states in the MES allows to obtain not only all pure fully entangled states but an arbitrary state (pure or mixed) of the same dimension via deterministic LOCC transformations. Moreover, we will present a six-qubit state that allows the local preparation of an arbitrary three-qubit state. Finally, we will discuss the differences that arise if one tries to extend our discussion to the  four-qubit case.

\begin{figure}[H]
\begin{center}
  \includegraphics[scale=0.45]{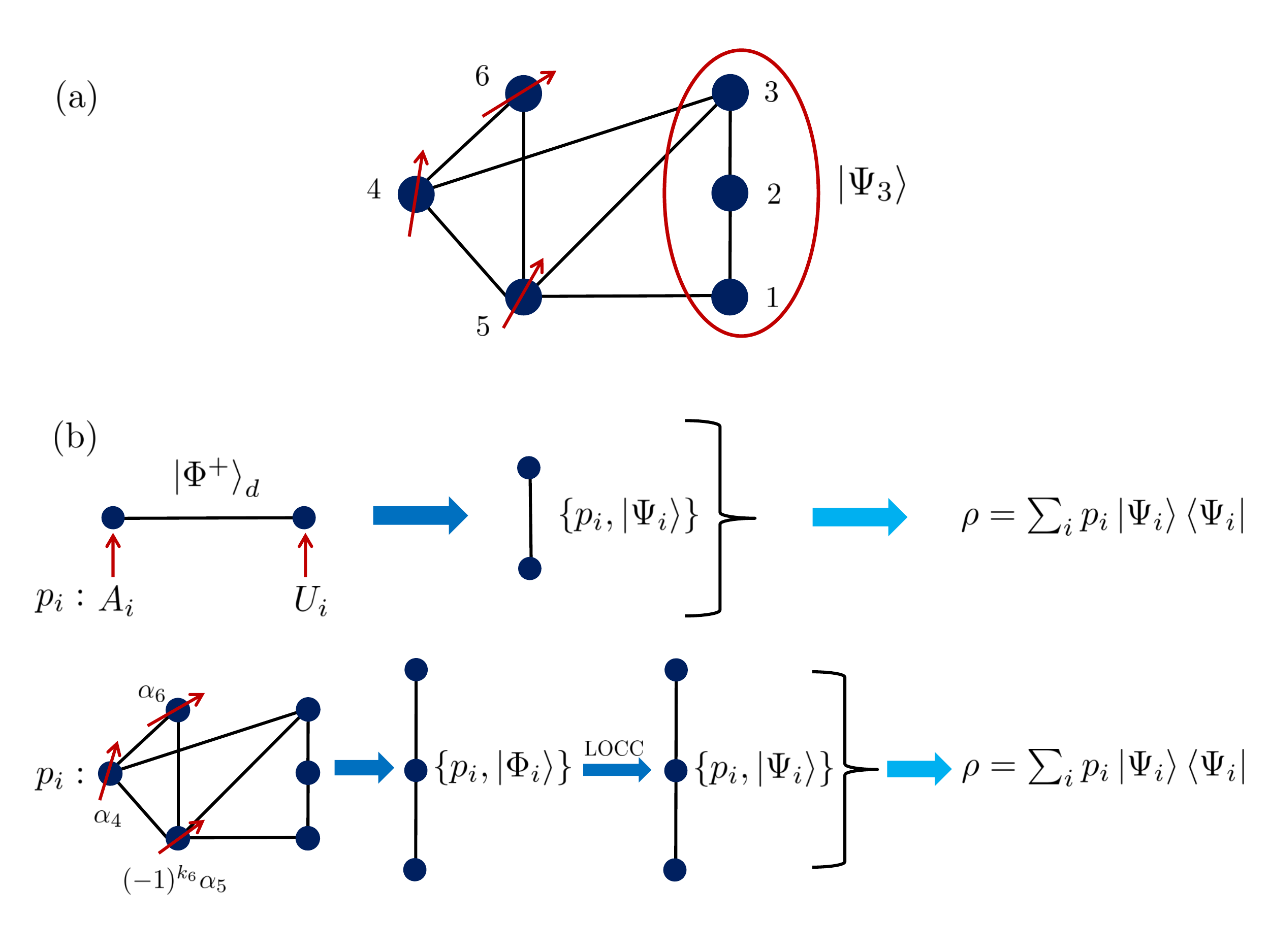}
  \caption{(a) This graphic shows the graph state that is LU-equivalent to the six-qubit state in Eq. (\ref{sixresource}). (b) In this graphic we indicate how any pure or mixed bipartite (three-qubit) state can be obtained via deterministic LOCC transformations from $\ket{\Phi^+}_d$ (the six-qubit state in Eq. (\ref{sixresource})) respectively.}
\label{repstate}
\end{center}
\end{figure}

In the bipartite case $MES_2$ contains a single state, namely $\ket{\Phi^+}_d$.   Interestingly, this state is not only a resource to prepare an arbitrary pure states but also to obtain any mixed bipartite state, $\rho=\sum_i p_i\ket{\Psi_i}\bra{\Psi_i}$. The protocol that allows to achieve this task is the following. With probability $p_i$ apply the LOCC protocol that deterministically transforms  $\ket{\Phi^+}_d$ into $\ket{\Psi_i}$ [see Fig. \ref{repstate} (b)]. Thus,  $\ket{\Phi^+}_d$ is not only the most useful state among pure states but also among mixed bipartite states.

In the three qubit case the MES, $MES_3$, contains infinitely many states. Having simultaneous access to all of the states in $MES_3$ allows to obtain any pure or mixed three-qubit state deterministically via LOCC as we will show in the following. It is clear from the definition of the MES that any pure fully entangled three-qubit state can be obtained via LOCC from one of the states in $MES_3$. Moreover, any bipartite entangled three-qubit state can be obtained from the GHZ-state by performing  a projective measurement in the $\sigma_x$-basis on the appropriate qubit (to obtain up to LUs $\ket{0}\ket{\Phi^+}_2$) and then applying the corresponding LOCC protocol to reach the desired state. It is clear that any product state can be reached from an arbitrary state in $MES_3$ via projective measurements. Thus, having access to all of the states in $MES_3$ allows to prepare any pure three-qubit state. Interestingly, it also allows to obtain any mixed three-qubit state as we will show in the following. In order to obtain a mixed three-qubit state, $\rho=\sum_i p_i\ket{\Psi_i}\bra{\Psi_i}$, choose with probability $p_i$ the state in $MES_3$ that allows to reach $\ket{\Psi_i}$ deterministically and apply the corresponding LOCC protocol to obtain $\ket{\Psi_i}$. Thus, being able to prepare any state in $MES_3$ and to perform LOCC transformations does not only allow to obtain any pure but also any mixed three-qubit state deterministically. This implies that  $MES_3$ contains also the most useful states in the mixed state case. 

Interestingly, there exists also a single six-qubit state that allows to obtain an arbitrary three-qubit state (pure or mixed) via deterministic LOCC transformations as we will explain in the following. Thus, this six-qubit state is a resource for three-qubit state preparation.
It has been introduced as a resource state in a different context, namely Remote Entanglement Preparation (REP) \cite{rep}. In REP the aim is that party A (which can be split into several spatially separated parties) is able to deterministically provide the spatially separated parties $B_i$ with arbitary multipartite entanglement. In the three qubit case it has been shown that a specific eight-qubit state has to be shared between the parties to provide the parties $B_i$ with an arbitary state up to LUs.
The scenario that we are interested in here is the deterministic preparation of an arbitrary state in $MES_3$. The corresponding resource has been shown to be the following six-qubit stabilizer state \cite{rep} [see Fig. \ref{repstate} (a)]

\bea\label{sixresource} \ket{\Phi_3}&=Z_{3}(\frac{\pi}{4})H_{3}Z_2(\frac{\pi}{2})Z_{6}(-\frac{\pi}{4})Z_{5}(-\frac{\pi}{4})Z_{4}(-\frac{\pi}{4})H_{1}S_{46}S_{56}S_{45}S_{15}S_{35}
S_{34}S_{12}S_{23}\ket+^{\otimes 6},\eea
where $S_{ij} =\ket{0}\bra{0}\otimes\one+\ket{1}\bra{1}\otimes \sigma_z$ and here and in the following we denote by $Z_{i_1\ldots i_m}(\alpha)$ a phase gate $e^{i\alpha\sigma_z^{i_1}\otimes\ldots\otimes\sigma_z^{i_m}}$ acting  non-trivially  on parties $i_1\ldots i_m$.
By performing projective local measurement on the qubits $4, 5$ and $6$ one prepares the state for qubits $1,2$ and $3$. The choice of the measurement basis decides on the values of the parameters in the following decomposition \begin{eqnarray}\label{form3}\ket{\Psi_3}=Z_{13}(\alpha_4) Z_{12}(\alpha_5) (T_2 \otimes T_3)Z_{23}(\alpha_6)\ket+^{\otimes 3},\end{eqnarray} where $T_3=  e^{-i\frac{\pi}{4}\sigma_x }Z(-\frac{\pi}{4})H$ and $T_2=e^{i\frac{\pi}{4}\sigma_y } Z(\frac{\pi}{4}) H $.
Note that any state that can be given (up to LUs) in the decomposition of Eq. (\ref{3qubitdecomp}) can  be written up to LUs also in this form. \footnote{Note that also any product or biseparable state can be written up to LUs in the decomposition given in Eq.(\ref{form3}).}.
Thus, via choosing the measurement basis accordingly one can prepare any state in  $MES_3$ (up to LUs) on the qubits $1,2$ and $3$. Let us now present some details of the protocol.
In order to specify the parameters in Eq. (\ref{form3}) qubit $4, 5$ and $6$ have to be measured in the basis $\{\sigma_z^{k_i} Z(-\theta_i)\ket{+}\}_{k_i=0,1}$ with $\theta_i=\alpha_i$ for $i=4, 6$ and $\theta_5=\pm\alpha_5$ for $i=5$ where the sign depends on the outcome of the measurement on qubit $6$. Thus, the measurement on qubit $6$ that determines the parameter $\alpha_6$ has to be performed before the measurement on qubit $5$ and depending on the measurement outcome the basis of the measurement that determines $\alpha_5$ might have to be adjusted (for outcome $k_6=1$ one has to choose $-\alpha_5$ instead of $\alpha_5$). Moreover, the state  on the qubits $1,2$ and $3$ is prepared up to Pauli operators which depend on the measurement outcomes. More precisely, one obtains the following state $(\sigma_z^{k_4+k_5}\otimes\sigma_z^{k_5}\sigma_y^{k_6}\otimes\sigma_z^{k_4+k_6})\ket{\Psi_3}$. The origin of the importance of the order of the measurements, as well as of the fact that the state is prepared up to Pauli operators is that we used deterministic gate implementation in order to construct the resource state (for details see \cite{rep}). As the six-qubit state in Eq.(\ref{sixresource}) allows to obtain any state in $MES_3$, it is  a resource for the preparation of  any pure or mixed three-qubit state via deterministic LOCC transformations [see Fig. \ref{repstate} (b)]. 

Note that there exists a different six-qubit state, namely $\ket{\Phi^+}_2^{\otimes 3}$ which is shared among A (who possesses 4 qubits), B (1 qubit) and C (1 qubit), that allows to obtain any state in $MES_3$ via a projective measurement and application of LUs, i.e. by teleporting the corresponding state in $MES_3$ \cite{teleport}. In fact, any pure (or mixed) three-qubit state can be obtained from $\ket{\Phi^+}_2^{\otimes 3}$ by performing (with some probability) a corresponding measurement and applying depending on the measurement outcome some Pauli operators. The important difference to our scheme is that except for fully-separable states the projective measurement is non-local and therefore it does not correspond to a LOCC protocol on the 6 qubits. 

In the four-qubit case the 23-qubit state that is a resource state for REP  can be used to obtain any pure or mixed four-qubit state, as it allows to prepare, by performing local projective measurements, an arbitrary pure four-qubit state up to LUs \cite{rep}. Unfortunately, the number of qubits of the resource state for REP in the four qubit case can not be reduced by  requiring only the preparation of states in the MES. The reason for this is the following. The fact that $MES_4$ is of full measure implies that (independent of the chosen decomposition) the number of parameters that describe $MES_4$ and the number of parameters that describe an arbitrary four-qubit state up to LUs have to be the same.  By construction the number of qubits of the resource state depends on the number of parameters in some decomposition of specific form (see \cite{rep} for details).  

Moreover, there is another clear difference to the three-qubit case.
In the four-qubit case having access to all states in the MES does not allow to obtain all biseperable states via deterministic LOCC transformations. States of the form $\ket{\phi}\ket{\psi}$, where $\ket{\phi}$ and $\ket{\psi}$ are truly bipartite entangled, can not be obtained from some state in $MES_4$. This can be  seen as follows. Applying a LOCC protocol to a state $\ket{\Psi}\in MES_4$ leads in any branch of the protocol to a state $M_k^1\otimes M_k^2\otimes M_k^3\otimes M_k^4\ket{\Psi}$ where $M_k^i$ are 2 x 2 matrices. Clearly, states of the form $\ket{\phi}\ket{\psi}$, where $\ket{\phi}$ and $\ket{\psi}$ are truly bipartite entangled, can not be obtained from  $\ket{\Psi}$ via a local projective measurement, i.e. all the matrices $M_k^i$ have to have rank 2 and are therefore invertible. This implies that $\ket{\Psi}$ and  $M_k^1\otimes M_k^2\otimes M_k^3\otimes M_k^4\ket{\Psi}$ have to be in the same SLOCC class, which is not the case for  $\ket{\Psi}\in MES_4$ and $\ket{\phi}\ket{\psi}$, where $\ket{\phi}$ and $\ket{\psi}$ are truly bipartite entangled. Hence, having access to all state in $MES_4$ and being able to perform any LOCC protocol does not allow to obtain all biseparable states and therefore not all four-qubit states. As the 23-qubit state that is a resource state for REP allows to obtain any four-qubit state (including also all biseparable four-qubit states) by performing the corresponding local measurements this state allows to prepare any mixed or pure four-qubit state via LOCC. 

In summary, we have shown that in the bipartite- and three qubit case the states that are contained in the MES are the most useful ones among all states (pure or mixed) of the same dimension. We presented a single six-qubit state from which any three-qubit state (pure or mixed) can be obtained deterministically by applying the corresponding LOCC protocol. Moreover, there exists a 23-qubit state that allows to obtain any pure or mixed four-qubit state via deterministic LOCC transformations.
\section{Summary and outlook}

The characterization of the possible LOCC transformations among quantum states is a very relevant problem, both in the foundations of entanglement theory and to identify the most useful states in order to look for new applications of quantum information. Although this is in general a very hard task, we have shown that it is in principle possible to address this question using the tools that characterize the conversions under the mathematically more tractable set of SEP operations. Following this idea, we have characterized the MES of truly entangled three-qubit states and generic four-qubit states. Interestingly, it turned out that there is a subset of states of measure zero which is clearly more relevant for state manipulation and these families should therefore be good candidates to look for practical applications. $MES_n$ allows to obtain by LOCC any possible $n$-qubit truly entangled pure state. However, we have shown that the power of $MES_3$ extends to the preparation of all possible (pure or mixed) tripartite qubit states but that this is not the case for $MES_4$. As a complement, we have considered LOCC protocols that allow to obtain all few-body states from a single state of more parties. We have provided a specific six-qubit (23-qubit) state that allows to prepare any form of three-qubit (four-qubit) entanglement.

Our results open the door to a systematic and general study of LOCC manipulation in multipartite systems. In the future we will investigate extensions to systems of higher dimension and/or more parties. Moreover, we will study the case of more copies and also the more general case of approximate LOCC transformations in which, contrary to exact LOCC, one allows for a certain error in the output state of the protocol. Furthermore, the investigation of possible LOCC transformations will allow, similarly to the bipartite case, to introduce operational entanglement measures.

This research was funded by the Austrian Science Fund (FWF): Y535-N16. JdV further acknowledges the Spanish MINECO through grants MTM2014-54692 and MTM2014-54240-P and the CAM regional research consortium QUITEMAD+CM S2013/ICE-2801. 

\end{document}